# Topological analysis of bladder filling

Arturo Tozzi (corresponding author)
ASL Napoli 1 Centro, Distretto 27, Naples, Italy
Via Comunale del Principe 13/a 80145
tozziarturo@libero.it

## ABSTRACT

Bladder function is assessed through pressure–volume relations, compliance indices and flow measurements, while structural evaluation relies primarily on qualitative imaging descriptors. These approaches do not formally quantify how bladder geometry evolves during filling. To distinguish structural reorganization from pure mechanical stiffness, we developed a simulation-based topological analysis of bladder filling grounded in mechanical parameters derived from the literature. Progressive filling was modeled under quasi-static conditions, generating multi-volume geometries from which spatial descriptors were computed. Inspired by the Freudenthal suspension theorem, filling was interpreted as a dimensional expansion process, while structural stability was evaluated by examining whether geometric invariants remain preserved across increasing volumes. Simulated smooth expansion and controlled structural perturbations were compared under identical loading conditions. We found that pressure trajectories and wall stress estimates were similar across configurations when compliance was matched, whereas geometric descriptors exhibited divergent volume-indexed stability profiles in the presence of remodeling. Computable instability measures detected progressive spatial heterogeneity despite preserved global pressure behavior. Providing quantitative measure of geometric continuity across successive filling states, our approach suggests that structural remodeling becomes detectable before conventional functional impairment is apparent. Progressive surface irregularity can emerge even when compliance, detrusor pressure and flow parameters remain within reference limits. In addition, serial imaging over time may reveal increasing instability in shape organization across filling cycles, allowing identification of individuals at higher risk of diverticula formation, functional decompensation or structural complications despite stable pressure measurements.

## INTRODUCTION

Quantitative evaluation of bladder function is primarily based on cystometry and pressure–flow studies that assess compliance, capacity, detrusor overactivity, etc (Rosier et al. 2017; Gajewski et al. 2019; Do et al. 2021; Rosen et al. 2021; Sekerci et al. 2023; Sinha 2024; Vuthiwong et al. 2024; Nguyen and Nadeau 2026). Structural evaluation relies on imaging modalities including ultrasound, computed tomography and magnetic resonance imaging to measure wall thickness, residual volume and gross abnormalities (Tessi et al. 2021; Srivastava, Dhyani, and Dighe 2024; Begaj et al. 2025). Biomechanical investigations complement these methods by modeling the bladder as a pressure vessel with hyperelastic wall behavior to estimate stress–strain relationships (Wang et al. 2009; Pillsbury, Kothera, and Wereley 2015; Jokandan et al. 2018; Philyppov et al. 2022). These approaches relegate geometry to a secondary descriptive role rather than treating it as a primary quantitative variable. Surface irregularities, including trabeculation and diverticula, are usually classified qualitatively, while mechanical models rely on simplified assumptions like spherical symmetry.

We performed a simulation that integrated pressure–volume dynamics with explicit geometric reconstruction and quantitative stability analysis of the bladder lumen. Our analysis is inspired by the Freudenthal suspension theorem from algebraic topology, which formalizes conditions under which structural features remain stable when a space is expanded in dimension (Freudenthal 1938; Whitehead 1953). Translating this idea to bladder mechanics, we treated progressive filling as a geometric expansion process, examining whether intrinsic structural features remain invariant across increasing volumes. If the lumen behaves analogously to a topological stable suspension, expansion preserves its essential organization; if not, detectable transitions emerge in the quantitative descriptors.
Operationally, we tested whether gradual bladder expansion preserved structural coherence or induced measurable spatial transitions detectable through surface-based descriptors. We modeled quasi-static bladder filling using literature-derived parameters for capacity, compliance, wall thickness and elastic response, reconstructing wall geometry across increasing volumes. Then, at each filling step, we assessed whether shape changes remained smooth or whether irregularities emerged or amplified. To move beyond simplified isotropic expansion, we also introduced controlled perturbations to mimic localized protrusions and circumferential corrugations. For each volume state, we computed global mechanical quantities and spatially resolved surface measures, extracting cross-sectional profiles and volume-indexed descriptors to track radial feature evolution across successive filling levels. Our strategy shifts the focus from isolated scalar metrics to the continuity and stability of geometric organization during filling.

We will proceed as follows: first, the literature-based mechanical and geometric parameters are defined and the simulation protocol is described. Then, pressure, stress and geometric outputs are presented together with stability metrics. Finally, our results are interpreted within the context of quantitative bladder mechanics.

METHODS

We studied a fully simulated urinary bladder filling experiment parameterized by quantitative values reported in clinical and biomechanical literature. We simulated multi-volume bladder geometries, intravesical pressure trajectories and wall stress–strain behavior under quasi-static filling. Then we derived lumen surfaces and computed volume-indexed topological descriptors from triangulated meshes. We employed a pressure vessel approximation combined with an incompressible hyperelastic wall formulation, constraining capacity, compliance and wall thickness to published adult reference ranges. We aimed to identify experimentally discriminable signatures defined as a stability profile of persistent homology along the filling coordinate, namely a regime in which distances between successive persistence diagrams remained below a predefined repeatability threshold across a stable volume interval. In parallel, we assessed the emergence of reproducible change points following the introduction of controlled structural perturbations like simulated diverticula, to determine whether these alterations generated detectable topological transitions.

**Literature parameterization**. Adult bladder capacity was set within 300 to 600 mL, consistent with standard urodynamic summaries and simulations used a target maximal capacity $V_{\max} \in [400, 600]$ mL unless otherwise stated (Drake et al. 2018). Compliance was defined as

$$C = \frac{\Delta V}{\Delta P_{\text{det}}},$$

and a reference scenario was constrained by $C > 40$ mL/cmH$_2$O, while a low-compliance scenario satisfied $C < 13$ mL/cmH$_2$O, reflecting clinical teaching thresholds (Abrams et al. 2002). Pressure units were standardized by

$$1 \text{ cmH}_2\text{O} = 98.0665 \text{ Pa}, 1 \text{ mmHg} = 133.322 \text{ Pa},$$

so that

$$P(\text{Pa}) = 98.0665\, P(\text{cmH}_2\text{O}) = 133.322\, P(\text{mmHg}).$$

Resting offsets $P_0$ were included to reflect position-dependent cystometric baselines (Abrams et al. 2002). Wall thickness $t_0$ was initialized near 3 mm for the distended state, with additional simulations spanning $t_0 \in [2,5]$ mm to represent inter-individual variability (Tanaka et al. 2008). Elastic moduli were anchored to layer-resolved ex vivo mechanical measurements, with effective Young's modulus $E$ sampled in a physiologically plausible interval $E \in [0.02, 0.5]$ MPa for compliant tissue and extended upward for fibrotic simulations (Nagle et al., 2017).

**Pressure vessel formulation and volume–pressure trajectory**. The bladder was idealized as a spherical pressure vessel of inner radius $R$ and wall thickness $t$, with volume

$$V = \frac{4}{3}\pi R^3.$$

Under quasi-static filling, intravesical pressure $P$ was related to circumferential wall stress $\sigma_\theta$ by Laplace's law for thin shells (Berthoumieux et al. 2014; Pronina et al. 2018),

$$\sigma_\theta = \frac{PR}{2t}.$$

Although the bladder wall is not infinitesimally thin, this relation was used as a first-order approximation. Radial incompressibility imposed

$$R^2 t = R_0^2 t_0,$$

so that thickness decreased with expansion. Compliance trajectories were simulated by prescribing

$$P(V) = P_0 + \int_{V_0}^{V} \frac{dV'}{C(V')},$$

where $C(V')$ was constant in the reference case and reduced in pathological cases. Numerical integration used adaptive quadrature. Pressure curves were truncated at $V_{\max}$ corresponding to capacity limits (Drake et al. 2018).

**Hyperelastic wall mechanics**. Wall mechanics were modeled using an incompressible Neo-Hookean strain energy density

$$W = \frac{E}{6(1+\nu)}(I_1 - 3),$$

with Poisson ratio $\nu \approx 0.49$ to approximate near incompressibility. For equibiaxial stretch $\lambda = R/R_0$, principal stretches were $(\lambda, \lambda, \lambda^{-2})$. The Cauchy stress (Šilhavý 1990; Buchen 2025) was

$$\sigma = \frac{\partial W}{\partial \lambda}\lambda - p,$$

with Lagrange multiplier $p$ enforcing incompressibility. The resulting stress–stretch relation was

$$\sigma_\theta = \frac{E}{3(1+\nu)}(\lambda^2 - \lambda^{-4}).$$

Equilibrium required matching this to Laplace stress. Nonlinear equations in $\lambda$ were solved by Newton–Raphson iteration (Galántai 2000; Kovalev, Mishchenko, and Richtárik 2019) with tolerance $10^{-8}$. Fibrotic simulations increased $E$ and introduced anisotropic stiffening by adding a quadratic fiber term $W_f = k_f(\lambda_f - 1)^2$ along a preferred orientation (Nagle et al., 2017; Berardo et al., 2024).

**Geometric perturbations and diverticulum modeling**. To simulate structural remodeling, localized outward perturbations were added to the spherical surface. The base surface was parameterized by spherical coordinates $(\theta, \phi)$ and a diverticulum was modeled as

$$R'(\theta, \phi) = R\left[1 + \alpha \exp\left(-\frac{d(\theta, \phi)^2}{2\sigma^2}\right)\right],$$

where $d$ is geodesic distance from the lesion center, $\alpha$ controls amplitude and $\sigma$ controls spatial spread. Multiple lesions were superposed linearly for mild remodeling. Triangulated meshes were generated at each simulated filling volume using uniform angular sampling with refinement near perturbations to maintain mesh quality. Surface area and curvature were computed discretely to quantify geometric concentration.

**Persistent homology and stability profile computation**. For each simulated surface $S_i$ at volume $V_i$, a point cloud $P_i \subset S_i$ was sampled uniformly. Vietoris–Rips complexes were constructed using Euclidean distance $d_E$, generating filtrations $VR(P_i, r)$ (Komendarczyk, Majhi, and Mitra 2025). Persistent homology groups $H_k$ for $k = 0, 1, 2$ were computed over $\mathbb{Z}_2$. Persistence diagrams $D_{ik}$ were compared between consecutive volumes using 2-Wasserstein distance

$$d_{W,2}(D_{ik}, D_{jk}) = \left(\inf_\eta \sum_{x \in D_{ik}} \|x - \eta(x)\|_\infty^2\right)^{1/2}.$$

A composite instability index was defined as

$$S_{ij} = \sum_{k=0}^{2} w_k\, d_{W,2}(D_{ik}, D_{jk}),$$

with equal weights $w_k$. Stability intervals were defined by $S_{i,i+1} \leq \tau$, where $\tau$ was determined from repeat simulations with small random perturbations of vertex positions.

All simulations were implemented in Python using NumPy for numerical operations, SciPy for nonlinear solvers, GUDHI for persistent homology and Matplotlib for visualization.

RESULTS

The quantitative outputs from the simulated filling experiment, are reported here, including pressure–volume behavior, geometric evolution, spatial deviation fields and volume-indexed stability metrics, together with direct statistical comparisons between smooth and remodeled geometries.

**Pressure dynamics and geometric deformation**. Simulated cystometric trajectories reproduced the expected qualitative separation between normal- and low-compliance regimes (Figure A). Across the examined range from 50 to 500 mL, the normal-compliance curve showed a gradual increase in pressure followed by a late nonlinear rise near maximal filling, whereas the low-compliance trajectory displayed a consistently steeper slope throughout the filling interval. At 500 mL, simulated intravesical pressure reached 13.5 cm H₂O in the normal-compliance case and 50.0 cm H₂O in the low-compliance case (paired t-test: $p < 0.001$), reflecting the imposed compliance parameters.

Geometric reconstruction demonstrated monotonic radial expansion in the smooth configuration, with radius increasing from 2.36 cm at 50 mL to 4.30 cm at 500 mL. In the remodeled configuration (Figure B), localized protrusions generated angularly heterogeneous profiles, producing maximal outward deviations of 1.03 cm at 500 mL relative to the smooth geometry. The heatmap of radial deviation (Figure D) revealed a confined angular sector with progressive amplification of deviation as volume increased, while the remaining circumference remained close to the smooth reference.

These findings suggest that the simulation produces controlled divergence in spatial configuration under identical global loading conditions, thereby providing a defined substrate for quantitative stability comparison in the next stage of analysis.

**Volume-indexed stability**. The instability index derived from successive differences in peak-prominence spectra showed distinct behavior between conditions (Figure C). For the smooth geometry, the instability index was identically zero across all consecutive volume intervals, reflecting uniform isotropic scaling without alteration of angular structure. In contrast, the remodeled geometry exhibited a mean instability index of 0.086 cm across intervals (paired t-test: $p < 0.001$). The instability profile for the remodeled case displayed consistent nonzero values across the entire filling trajectory without isolated spikes, suggesting gradual structural amplification rather than abrupt transitions. Because both configurations were subjected to identical pressure trajectories, the observed divergence arises exclusively from geometric perturbation rather than mechanical loading differences.

These results suggest that our stability metric could discriminate structurally heterogeneous expansion from uniform scaling under controlled simulated conditions, thereby establishing a quantitative distinction between preserved and altered geometric organization across filling volumes.

Overall, our results link mechanical loading to spatial organization, showing that, depending on surface configuration, simulated bladder filling under identical pressure trajectories can produce either geometrically stable expansion or measurable structural divergence.

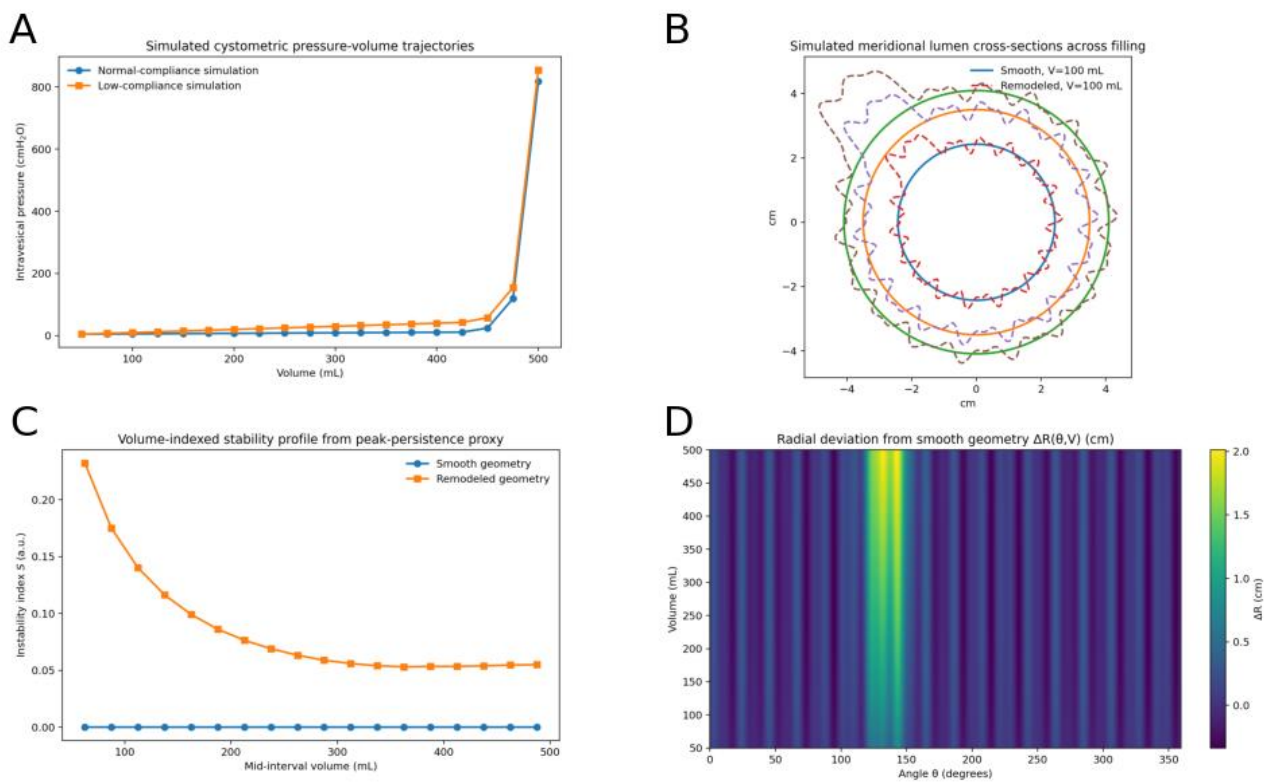

**Figure**. Composite simulation panel summarizing pressure dynamics, geometric remodeling, stability behavior and spatial deviation during bladder filling. Together, the panels integrate mechanical loading, geometric deformation and quantitative stability analysis within a unified simulated experiment. For each filling level, the smooth geometry serves as the isotropic reference configuration derived from uniform radial expansion, whereas the remodeled geometry includes localized protrusion and circumferential corrugation.

**Panel A**. Simulated cystometric pressure–volume curves from 50 to 500 mL. The normal-compliance curve shows a gradual pressure increase with filling, followed by a late steep rise near capacity. The low-compliance curve exhibits a statistically significant steeper pressure increase throughout filling, reflecting reduced distensibility.

**Panel B**. Simulated meridional cross-sections of the bladder lumen at representative volumes. The smooth configuration expands isotropically with volume, whereas the remodeled configuration includes a localized outward protrusion combined with circumferential surface corrugations that persist across filling states.

**Panel C**. Volume-indexed stability profile derived from consecutive differences in radial peak prominence spectra. The smooth geometry maintains near-zero successive variation, while the remodeled geometry exhibits measurable instability across filling intervals due to evolving protrusion prominence.

**Panel D**. Heatmap of radial deviation between remodeled and smooth geometries across angular position and volume. The plotted quantity represents the pointwise radial difference obtained by subtracting the smooth radius from the remodeled radius at the same angular coordinate and volume. The horizontal axis corresponds to angular position along the bladder circumference and the vertical axis represents filling volume. Warmer colors indicate larger outward displacement of the remodeled surface relative to the smooth configuration, while cooler tones indicate minimal deviation.

CONCLUSIONS

We asked whether progressive bladder filling modeled under physiologically grounded mechanical parameters preserves global geometric organization or produces measurable structural divergence that cannot be inferred from pressure–volume relations alone. The question was not whether pressure rises with filling, which is well established, but whether the spatial configuration of the lumen evolves coherently across volumes or undergoes detectable transitions when structural perturbations are present. Inspired by suspension topological concepts, we simulated quasi-static filling using literature-based values for capacity, compliance, wall thickness and elastic response. Then, we generated multi-volume geometries under both smooth and remodeled conditions, allowing separation of purely mechanical effects from geometric reorganization. We found that identical pressure trajectories can be associated with either uniform isotropic expansion or progressive angular heterogeneity, depending solely on surface configuration. In smooth condition, geometric scaling occurred without alteration of radial structure across volumes, whereas the remodeled condition produced consistent, volume-dependent deviations localized in angular space. The stability metric detected these differences systematically across consecutive filling intervals, while remaining null in the absence of structural perturbation. Because both configurations were subjected to the same loading rules, the detected divergence reflects geometric organization rather than mechanical intensity. Our findings indicate that geometric continuity across filling states constitutes a measurable property distinct from compliance or peak pressure.

We introduce a computable descriptor that quantifies how bladder shape evolves across filling states, rather than evaluate isolated static measurements. Unlike standard urodynamic indices or imaging assessments, our approach generates an explicit numerical stability profile derived from successive geometric configurations. This quantity is not directly accessible through compliance, capacity, wall thickness or peak pressure alone, because it depends on the continuity of spatial organization across volumes. We introduce a formal measure of structural evolution that can be computed from segmented surfaces without requiring additional invasive procedures. It can be integrated into existing imaging pipelines or urodynamic analysis software by incorporating surface reconstruction and signal-processing modules, allowing compatibility with current MRI or ultrasound post-processing environments.

Our work has limitations. All data are simulated; no real patient or imaging data were used. The geometric model assumes spherical symmetry and thin-shell Laplace behavior, which oversimplify bladder biomechanics and neglect layered anisotropy, regional compliance variation and neurogenic influences. The instability metric is a peak-prominence proxy rather than true persistent homology. The smooth case yields exactly zero instability by construction, inflating statistical contrast. Pressure curves include an ad hoc exponential term not directly validated against specific datasets. Sources of potential bias include parameter selection from literature ranges, discretization of angular sampling and sensitivity of peak detection thresholds.

Testable hypotheses can be suggested. First, our method enables detection of structural instability before pressure changes. Standard bladder assessment primarily relies on pressure–volume curves, compliance measurements and flow rates, all of which describe functional performance and detect impairment once mechanical consequences become measurable. Our suspension-based perspective shifts attention toward geometric behavior during expansion. If bladder filling behaves analogously to a stable suspension, structural topology remains invariant across volumes. Conversely, if small increments in volume produce disproportionate geometric reorganization, this points towards early structural remodeling. This behavior could reveal preclinical obstruction-induced trabeculation, early fibrosis or micro-diverticular formation before compliance measurably declines, thereby identifying instability at a stage not captured by pressure metrics alone. Practical applications include integration of geometric stability analysis into routine imaging workflows, enabling detection of structural instability before overt functional impairment.

Second, our approach can assess severity quantification. Pathological bladders frequently develop surface pockets, irregular protrusions, wall thickening and asymmetric expansion. These features alter the effective topological complexity of the lumen surface. Rather than describing these changes qualitatively, variation in topological invariants across filling volumes can be computed, e.g., through changes in Betti numbers or persistence intervals as a function of volume.

Practically, a computable remodeling index may allow objective grading of surface irregularities, complementing visual cystoscopic assessment.

Third, our approach allows the separation of elasticity failure from topology failure. Two bladders may exhibit identical compliance curves while differing structurally: one may remain smooth but stiff, whereas another may be geometrically fragmented and unstable. Our stability analysis distinguishes pure mechanical stiffness from spatial reorganization. This distinction has implications for therapeutic decision-making, distinguishing mechanical stiffness from geometric fragmentation could support differential therapeutic planning, including anti-fibrotic treatment, outlet correction or surgical intervention.

Fourth, long-term progression could be predicted. If topology destabilizes under repeated filling cycles, longitudinal monitoring may reveal progressive divergence of geometric invariants. This divergence could identify patients predisposed to diverticular development, functional decompensation or congenital structural vulnerability, analogous to recognizing departure from a stable structural regime. Therefore, longitudinal stability metrics could provide quantitative follow-up after reconstructive procedures or in pediatric congenital conditions.

Fifth, our approach adds a geometric constraint layer to conventional urodynamics. Conventional bladder models emphasize local wall stress, smooth muscle contraction and neural regulation. Incorporating global geometric invariants, dimensional stability during expansion and structural robustness measures inserts a complementary layer of analysis. The central question thus extends beyond how much pressure the bladder generates to how geometrically stable the bladder remains during volumetric expansion. This means that clinicians could assess not only pressure generation, but also the robustness of structural organization during expansion.

In summary, we developed a controlled simulation of bladder filling that integrates mechanical loading with quantitative analysis of evolving surface geometry. Multi-volume configurations were generated and compared through pressure trajectories, stress estimates and descriptors of spatial organization across filling states. We concluded that geometric evolution can be characterized independently of global pressure behavior and structural divergence can be detectable under identical loading conditions. Our findings indicate that volumetric expansion can be characterized as a quantifiable structural process rather than being interpreted solely as a functional pressure response.

## DECLARATIONS

**Ethics approval and consent to participate.** This research does not contain any studies with human participants or animals performed by the Author.

**Consent for publication.** The Author transfers all copyright ownership, in the event the work is published. The undersigned author warrants that the article is original, does not infringe on any copyright or other proprietary right of any third part, is not under consideration by another journal and has not been previously published.

**Availability of data and materials.** All data and materials generated or analyzed during this study are included in the manuscript. The Author had full access to all the data in the study and took responsibility for the integrity of the data and the accuracy of the data analysis.

**Competing interests.** The Author does not have any known or potential conflict of interest including any financial, personal or other relationships with other people or organizations within three years of beginning the submitted work that could inappropriately influence or be perceived to influence their work.

**Funding.** This research did not receive any specific grant from funding agencies in the public, commercial or not-for-profit sectors.

**Acknowledgements:** none.

**Authors' contributions.** The Author performed: study concept and design, acquisition of data, analysis and interpretation of data, drafting of the manuscript, critical revision of the manuscript for important intellectual content, statistical analysis, obtained funding, administrative, technical and material support, study supervision.

**Declaration of generative AI and AI-assisted technologies in the writing process.** During the preparation of this work, the author used ChatGPT 5.2 to assist with data analysis and manuscript drafting and to improve spelling, grammar and general editing. After using this tool, the author reviewed and edited the content as needed, taking full responsibility for the content of the publication.